\documentclass[pre,a4paper,onecolumn,noshowpacs,noshowkeys,nofootinbib,floatfix]{revtex4}
\usepackage{amsfonts}
\usepackage{amsmath}
\usepackage{amssymb}
\usepackage{graphicx}

\begin{document}

\title{Are all highly liquid securities within the same class?}
\author{S.~M.~Duarte~Queir\'{o}s}
\email[e-mail address: ]{sdqueiro@cbpf.br}
\affiliation{Centro Brasileiro de Pesquisas F\'{\i}sicas, 150, 22290-180, Rio de Janeiro
- RJ, Brazil}
\date{\today}

\begin{abstract}
In this manuscript we analyse the leading statistical properties of fluctuations of ($\log $) $3$-month US
Treasury bill quotation in the secondary market, namely: probability density function, autocorrelation, absolute
values autocorrelation, and absolute values persistency. We verify that this financial
instrument, in spite of its high liquidity, shows very peculiar properties. Particularly, we verify that
$\log $-fluctuations belong to the L\'{e}vy class of stochastic variables.
\end{abstract}

\maketitle

Financial markets have become a paradigmatic example of complexity and the
focus of plenty of work within physics. Specifically, several techniques,
mainly related to statistical physics (\textit{e.g.}, stochastic dynamics,
theory of critical phenomena or nonlinear systems), have been applied either
to reproduce or simply verify several properties, \textit{e.g.}, the
probability density functional form (PDF), or the autocorrelation function
(ACF) of financial observables~\cite{books,book2,book3}. The
systematic (asymptotic) power-law behaviour found for quantities such as
price/index fluctuations, or traded volume has been pointed out to be at the
helm of the multifractal character of financial time series~\cite%
{mandelbrot-fractals-scaling}, a feature that is also regular in
out-of-equilibrium systems~\cite{amit}. On the account of the background on
this type of phenomena, in which scale invariance also rules, it has come
out the endeavour to identify universality classes for financial markets
defined by the exponents that characterise their main statistical
properties. Explicitly, these classes indicate the
existence of a common behaviour for systems within the same class apart
their microscopic or specific details~\cite{stanley-colloqium}. On this way,
it has been suggested \cite{stanley-colloqium} that financial products like
securities with a very high level of liquidity (high trading activity) might
present similar characteristics. As an example, it has been shown that,
despite of the fact that in their essence stocks and commodities are
completely different financial instruments (securities), their (daily) price fluctuations
behave on a very similar way, \textit{i.e.}, they can be enclosed in the
same class~\cite{matia-amaral-bp}.

Within securities are also public debt bonds like United States (US) Treasury
bills~\cite{debt,bali}. The US Treasury bills (T-bills) are marketable bonds
issued by the US federal government and represent one of the debt financing
instruments used by the Treasury Department~\cite{t-bill}. T-bills are
classified as \textit{zero-coupon} bonds that are sold in the primary market
at a discount of the face value in order to present a positive yield to
maturity which can be $28$ ($1$ month), $91$ ($3$ months), or $182$ ($6$
months) days. In regard of this, they are considered to be the most
risk-free investment in the USA. This makes of T-bills an important and
heavily traded (\textit{i.e.}, highly liquid) financial instrument in the
secondary market where they are quoted on an annual percentage yield to
maturity.

In the sequel of this manuscript we study some of the main statistical
features of the $3$-month US T-bills traded on the secondary market. Our
time series, $\left\{ Q_{t}\right\} $, which is named $DTB3$ by the Federal
Reserve, is composed by $3$-month US T-bill daily prices and runs from the $%
4^{th}$ January $1954$ up to the $26^{th}$ February $2007$ in a total of $%
13866$ trading days~\cite{data-url}. Our choice for a maturity of $3$ months
is justified by the fact that it is the most used interest rate maturity in
derivative financial products like call-put options. To compare the
statistical properties of $DTB3$ daily $\log $-value fluctuations, $\tilde{r}%
_{t}=\ln Q_{t}-\ln Q_{t-1}$, we use the daily $\log $-index fluctuations, $%
\tilde{r}_{t}^{\prime }=\ln S_{t}-\ln S_{t-1}$, of $SP500$ time series, $%
\left\{ S_{t}\right\} $, which runs the same time interval as $DTB3$. Both
fluctuation time series, $\left\{ \tilde{r}_{t}\right\} $ and $\left\{ 
\tilde{r}_{t}^{\prime }\right\} $ have been subtracted of respective
averages, $\left\langle \tilde{r}_{t}^{(\prime )}\right\rangle $, and
normalised by standard deviation $\sigma _{\tilde{r}^{(\prime )}}$, \textit{%
i.e.}, $r_{t}^{(\prime )}=\left[ \tilde{r}_{t}^{(\prime )}-\left\langle 
\tilde{r}_{t}^{(\prime )}\right\rangle \right] /\sigma _{\tilde{r}^{(\prime
)}}$. (from here on the prime stands for $SP500$ quantities, and $x$ is used
in definitions to represent any observable upon analysis).

\medskip

\begin{figure*}[tbp]
\begin{center}
\includegraphics[width=0.66\columnwidth,angle=0]{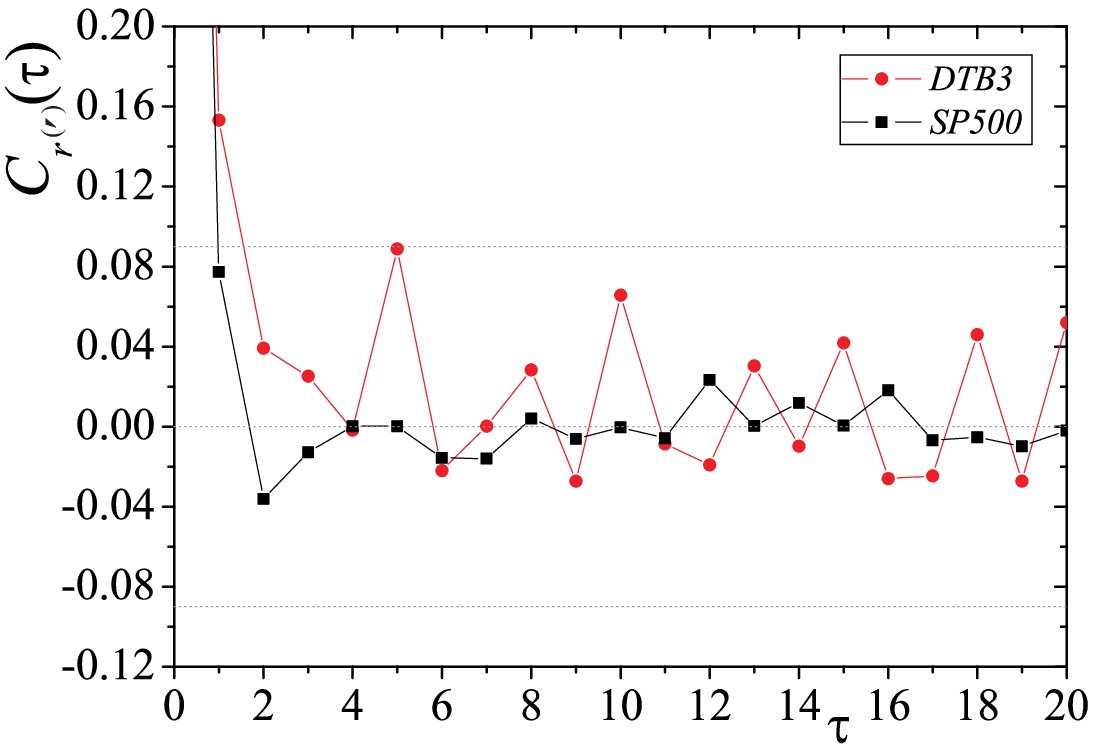} %
\includegraphics[width=0.66\columnwidth,angle=0]{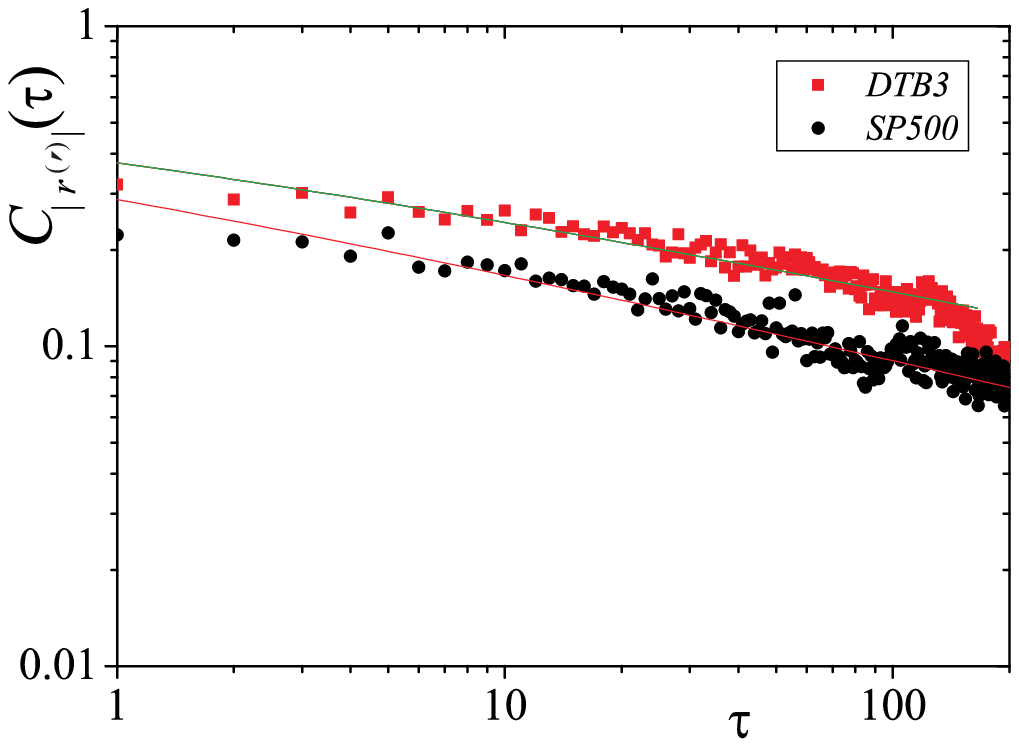}
\end{center}
\caption{(colour on-line) Autocorrelation function Eq.~\ref{cor-eq}, $C\left( \protect\tau \right) $
\textit{vs.} $\protect\tau $ of $r_{t}^{(\prime )}$ (left panel) and $%
\left\vert r_{t}^{(\prime )}\right\vert $ in a $\log $-$\log $ scale (right
panel). It is visible that $r_{t}$ is correlated for immediate correlations
and presents measurable correlations every multiple of $5$ days lag. The
autocorrelation of $\left\vert r_{t}^{(\prime )}\right\vert $ can be
described by Eq. (\protect\ref{q-exp}) with $q_{c}=4.7\pm 0.1$, $\mathcal{T}%
=0.45\pm 0.05$ ($\protect\chi ^{2}/n=2\times 10^{-4}$, $R^{2}=0.9$) for $DTB3
$, and $q_{c}^{\prime }=4.3\pm 0.1$, $\mathcal{T}^{\prime }=0.77\pm 0.05$ ($%
\protect\chi ^{2}/n=10^{-4}$, $R^{2}=0.9$) for $SP500$. The dashed line in left panel
represents three times the noise level bounds.}
\label{fig-1}
\end{figure*}

Moving ahead, we shall now analyse and compare primary and more usually
studied statistical features. Commencing with the analysis of ACF,
\begin{equation}
C_{x}\left( \tau \right) =\frac{\left\langle x_{t}\,x_{t+\tau }\right\rangle
-\left\langle x_{t}\,\right\rangle ^{2}}{\left\langle
x_{t}^{2}\,\right\rangle -\left\langle x_{t}\,\right\rangle ^{2}},
\label{cor-eq}
\end{equation}
we have verified a noteworthy difference between $\left\{ r_{t}\right\} $ and $%
\left\{ r_{t}^{\prime }\right\} $. Firstly, as depicted in Fig.~\ref{fig-1}, $C_{r}\left( 1\right) $ clearly exceeds three time noise level within 
which typical interday correlation values of $SP500$ and
other indices as well \cite{books} lay in. Additionally, correlation values greater
than noise level have been measured at least for lag $\tau =5,10$ days. We
attribute the origin of this feature to the fact that T-bills are weekly ($5$
trading days) sold at the primary market. Concerning the ACF of absolute
values, we have not observed any relevant differences. Both curves are
fairly described by (asymptotic power-law) $q_{c}^{(\prime )}$-exponential functions, 
\begin{equation}
C_{\left\vert r^{(\prime )}\right\vert }\left( \tau \right) =\,\left[
1-\left( 1-q_{c}^{(\prime )}\right) \,\mathcal{T}^{(\prime )}\mathcal{\,}%
\tau ^{2}\right] ^{\frac{1}{1-q_{c}^{(\prime )}}},  \label{q-exp}
\end{equation}
where $q_{c}^{(\prime )}$ gives the decaying exponent, and $\mathcal{T}%
^{(\prime )}$ characteristic parameter. The value $q_{c}=4.7\pm 0.1$ is not
far from $q_{c}^{\prime }=4.3\pm 0.1$, and both are in accordance with
previous values obtained for $SP500$~\cite{amaral-cps-pre} or $DJIA$
equities~\cite{smdq-canberra}. For $\mathcal{T} ^{(\prime )}$ we have obtained 
$\mathcal{T}=0.45\pm 0.05$, and $\mathcal{T}^{\prime }=0.77\pm 0.05$.

\begin{figure*}[tbp]
\begin{center}
\includegraphics[width=0.66\columnwidth,angle=0]{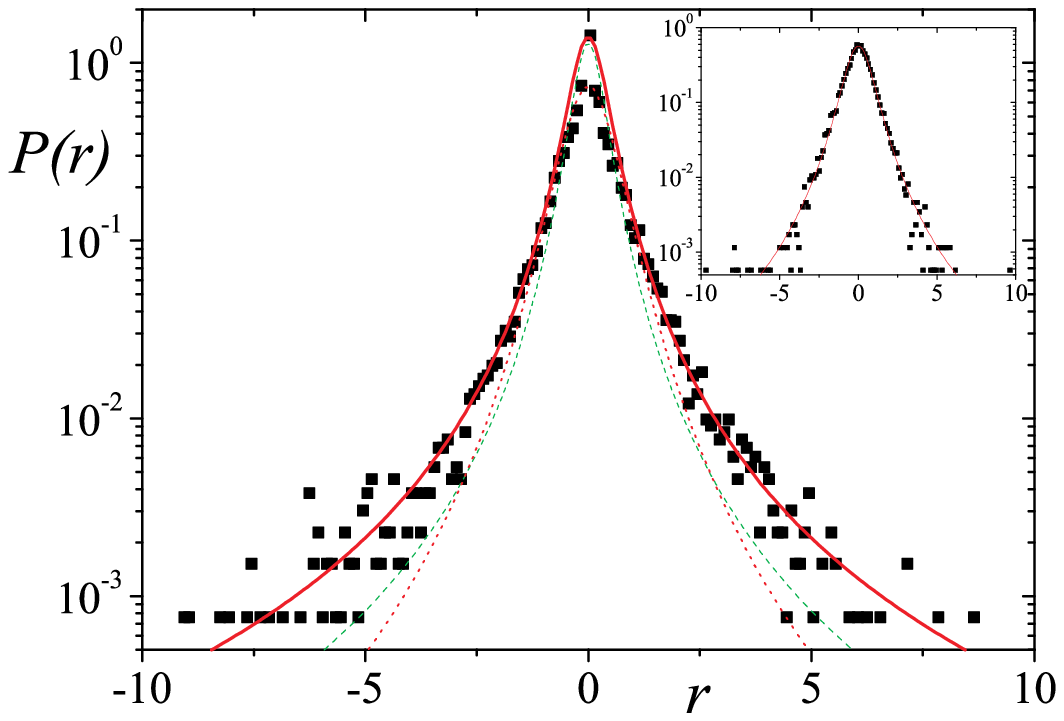} %
\includegraphics[width=0.66\columnwidth,angle=0]{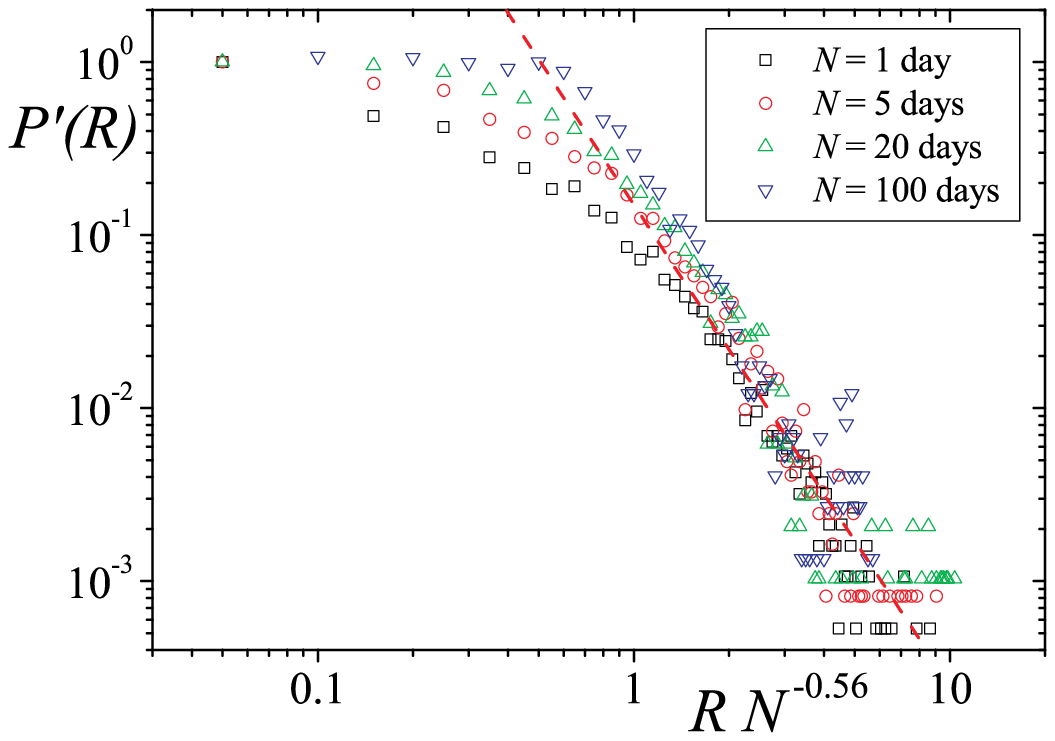}
\end{center}
\caption{(colour on-line) Left panel: PDF, $P\left( r\right) $ \textit{vs. }$%
r$ in $\log $-linear scale. Symbols are obtained from data and the full line
represents the best numerical adjustment for Eq. (\protect\ref{q-gauss}),
with $q=1.72\pm 0.02$ and $\mathcal{B}=5.9\pm 0.4$ ($R^{2}=0.96$ and $%
\protect\chi ^{2}/n=3\times 10^{-3}$). The dotted line is the best fit for $%
\mathcal{G}_{q}\left( r\right) $, but imposing $q=q^{\prime }=1.49$ as in $%
SP500$ case shown at the in-set ($\mathcal{B}^{\prime }=2.23\pm 0.09$, $%
R^{2}=0.99$, and $\protect\chi ^{2}/n=3\times 10^{-4}$ ). The dashed line
represents the best fit with $q=1.666$ ($\mathcal{B}=8.2\pm 0.6$) (on the
edge of finite variance). Right panel: $P^{\prime }\left( R_{N}\right) =%
\frac{P\left( R_{N}\right) }{P\left( 0\right) }N^{1/\protect\alpha .}$
\textit{vs. }$R_{N}\,N^{-1/\protect\alpha }$ for $N=1$, $5$, $20$, $100$
days in $\log $-$\log $ scale. The asymptotic collapse of the curves,
described by a tail exponent of $1+\protect\alpha =2\frac{1}{q-1}=2.77$ is
visible.}
\label{fig-2}
\end{figure*}

Stronger dissimilarity has appeared on the PDFs, which we have fitted for $q$%
-Gaussian distributions,%
\begin{equation}
\mathcal{G}_{q}\left( x\right) =\mathcal{A}\,\left[ 1-\left( 1-q\right) \,%
\mathcal{B\,}x^{2}\right] ^{\frac{1}{1-q}},\qquad \left( q<3\right) ,
\label{q-gauss}
\end{equation}%
where $\mathcal{A}$ is the normalisation, and $\mathcal{B}$ is related to
the \textquotedblleft width\textquotedblright\ of the distribution
determined by its $q$-generalised second order moment, $\sigma
_{q}^{\,2}=\int x^{2}\ \left[ P\left( x\right) \right] ^{q}\ dx/\int \left[
P\left( x\right) \right] ^{q}dx$, in the form, $\mathcal{B}=\left[ \sigma
_{q}^{\,2}\left( 3\,q-1\right) \right] ^{-1}$~\cite{tsallis-milan}. When $%
q<5/3$, standard deviation is finite and the equality $\mathcal{B}=\left[ \sigma
^{2}\left( 5-3\,q\right) \right] ^{-1}$ is also valid. For $q<3$,
Distribution (\ref{q-gauss}) emerges from optimising non-additive (Tsallis)
entropy upon appropriate constraints~\cite{ct}. In the limit $q\rightarrow 1$
the Gaussian distribution is obtained, $\mathcal{G}_{1}\left( x\right)
\equiv \mathcal{G}\left( x\right) $. Regardless both of the two fluctuations
are well described by Eq. (\ref{q-gauss}), the values of $q$ are
qualitatively quite different. Namely, we have obtained the best fit for $%
q=1.72\pm 0.02$ for $DTB3$, and $q^{\prime }=1.49\pm 0.01$ for $SP500$ (see
Fig.~\ref{fig-2}) \footnote{%
We have also used the Hill estimator to evaluate tail exponents. Due to
series length and error margins we cannot rely on the results obtained by
this method, although considering error margins they accord with $q^{(\prime
)}$ values.}. The latter is in accordance with prior analysis~\cite%
{books,book2,book3,amaral-cps-pre,q-arch}. Such a disparity has clear
implications on the attractor in probability space of each observable when
we consider the addition of fluctuations defining variable $%
R_{N,\,t}^{(\prime )}\equiv \sum_{i=0}^{N-1}r_{t+i}^{(\prime )}$. Since the
two signals are essentially uncorrelated, in the sense that ACF rapidly
attains at noise level, standard central limit theorems do apply~\cite%
{araujo}. In other words, for $SP500$, by reason of its entropic index $q$
is smaller than $\frac{5}{3}$, $\sigma ^{\prime }$ is finite, hence the convolution of
PDF $\log $-$SP500$ fluctuations leads to the Gaussian distribution, $%
\mathcal{G}\left( R_{N}^{\prime }\right) =\frac{1}{\sqrt{2\,\pi \,N\,
\left( \sigma ^{\prime } \right)
^{2}}}\exp \left[ -\frac{R_{N}^{\prime }}{2\,N\,\left( \sigma ^{\prime
}\right) ^{2}}\right] $ (for $N\rightarrow \infty $, and since the daily
time series has been normalised upon a finite series, $\sigma ^{\prime } \approx 1$).
Conversely, the entropic index for $DTB3$ is greater than $\frac{5}{3}$,
which makes $\sigma $ actually incommensurable. Thus, according to the L\'{e}%
vy-Gnedenko central limit theorem~\cite{araujo}, the attracting distribution
(for $N\rightarrow \infty $) is an $\alpha $-stable distribution,%
\begin{equation}
\mathcal{L}_{\alpha }\left( R_{N}\right) =\frac{1}{2\,\pi }\int_{-\infty
}^{+\infty }\exp \left[ -i\,k\,R_{N}-a\left\vert k\right\vert ^{\alpha }%
\right] dk,  \label{levy}
\end{equation}%
with $\alpha =\left( 3-q \right) /\left( q -1\right) $,
which follows, for large $N$, the scaling law $\mathcal{L}_{\alpha }\left(
R_{N}\right) =N^{-1/\alpha }\mathcal{L}_{\alpha }\left( \frac{R_{N}}{%
N^{1/\alpha }}\right) $, and $\mathcal{L}_{\alpha }\left( R_{N}\right) \sim
R_{N}^{-\alpha -1}$. As it is visible in Fig.~\ref{fig-2}, the PDFs of
properly scaled $R_{N}$ variables obtained from $r\left( t\right) $ signal
asymptotically collapse exhibiting a tail described by $\alpha \approx 1.77$%
, as it happens for variables whose attractor is a $\alpha $-stable
distribution (see Ref.~\cite{catania}). This constitutes, in our point of
view, a substantial difference between $3$-month T-bill daily fluctuations
and other financial fluctuations, by the fact that it represents a drastic
change of the attractor.

\begin{figure}[tbp]
\begin{center}
\includegraphics[width=0.66\columnwidth,angle=0]{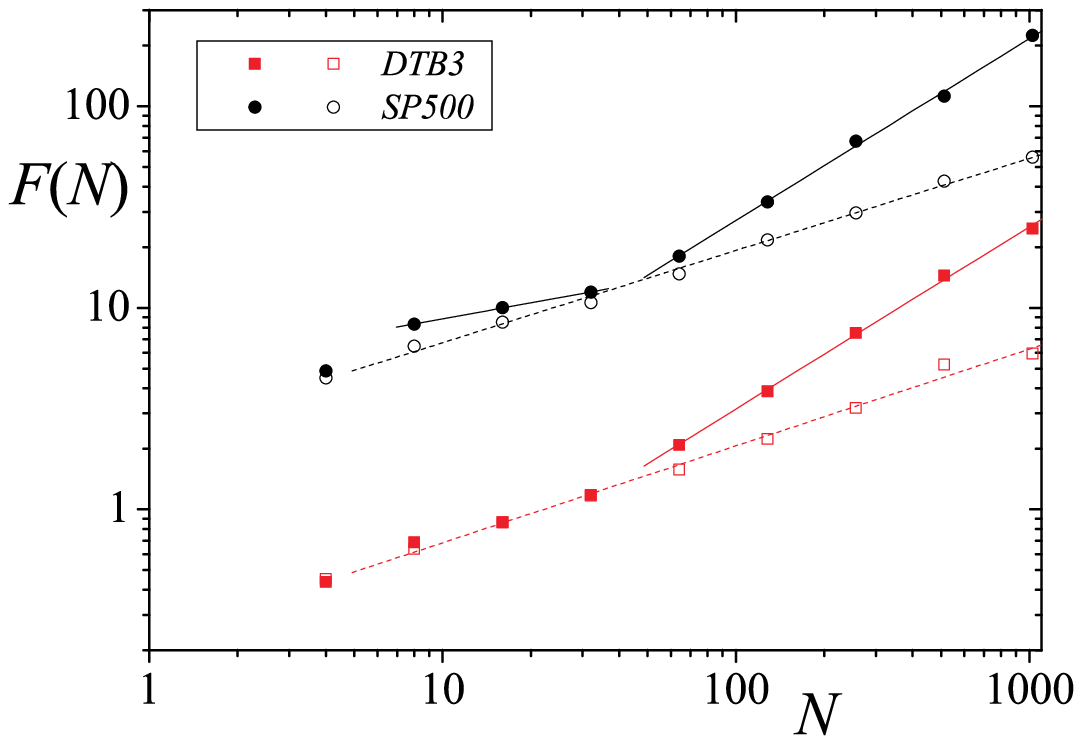} %
\end{center}
\caption{(colour on-line) Left panel: Root-mean-square deviation $F\left(
N\right) $ vs. $N$ for integrated absolute fluctuations time series of $DTB3$
(squares), and $SP500$ (circles). The full symbols are from the ordered
series and empty symbols from shuffled signals. For large $N$ we have
measured a Hurst exponent of $0.90\pm 0.02$ ($DTB3$), and $0.90\pm 0.03$ ($%
SP500$). For small $N$, T-bills absolute fluctuations $\log $-fluctuations presents
a behaviour similar to white noise while $SP500$ exhibits a antipersistent
behaviour.}
\label{fig-3}
\end{figure}

Within a macroscopic framework, the long-lasting form of the absolute price
fluctuations ACF has been held responsible for the non-Gaussian behaviour of
financial securities fluctuations~\cite{long-volatility,smdq-qf}. To further analyse
the persistency of absolute fluctuations, we have applied the $DFA$ method~to assess
the Hurst exponent, $H$,~of $\left\vert r_{t}^{(\prime )}\right\vert $ time series 
and shuffled $\left\{ \left\vert r_{t}^{(\prime )}\right\vert \right\} $
(procedure presented in \cite{dfa}). The results are exhibited in Fig.~\ref{fig-3}, where $N
$ represents the lenght of the time series \cite{dfa}. For $N>40$ we have
verified that $DTB3$ presents a strong persistent behaviour as $SP500$ does
with $H=0.90\pm 0.02$ and $H^{\prime }=0.90\pm 0.03$. For $N<40$ we verify a
crossover, but this time index and T-bill fall apart with $H=0.50\pm 0.02$
(like a Brownian motion) and a specious $H^{\prime }=0.27$
(antipersistency). It is known that the presence of spikes and locality on persistency 
might introduce spurious features on DFA analysis~\cite%
{dfa-bad,dfa-bad1,dimatteo} of persistent signals leading to $H<1/2\ $values
for small $N$. We attribute to this fact the emergence of $H\leq 1/2$ values.

\begin{figure*}[tbp]
\begin{center}
\includegraphics[width=0.48\columnwidth,angle=0]{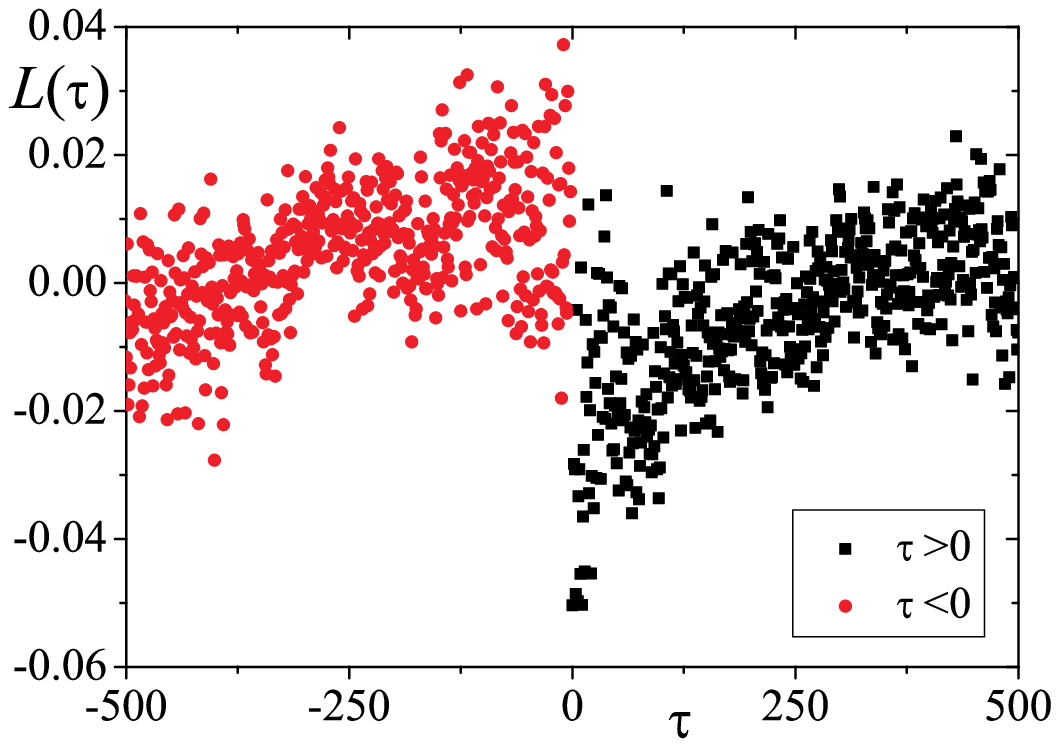} %
\includegraphics[width=0.48\columnwidth,angle=0]{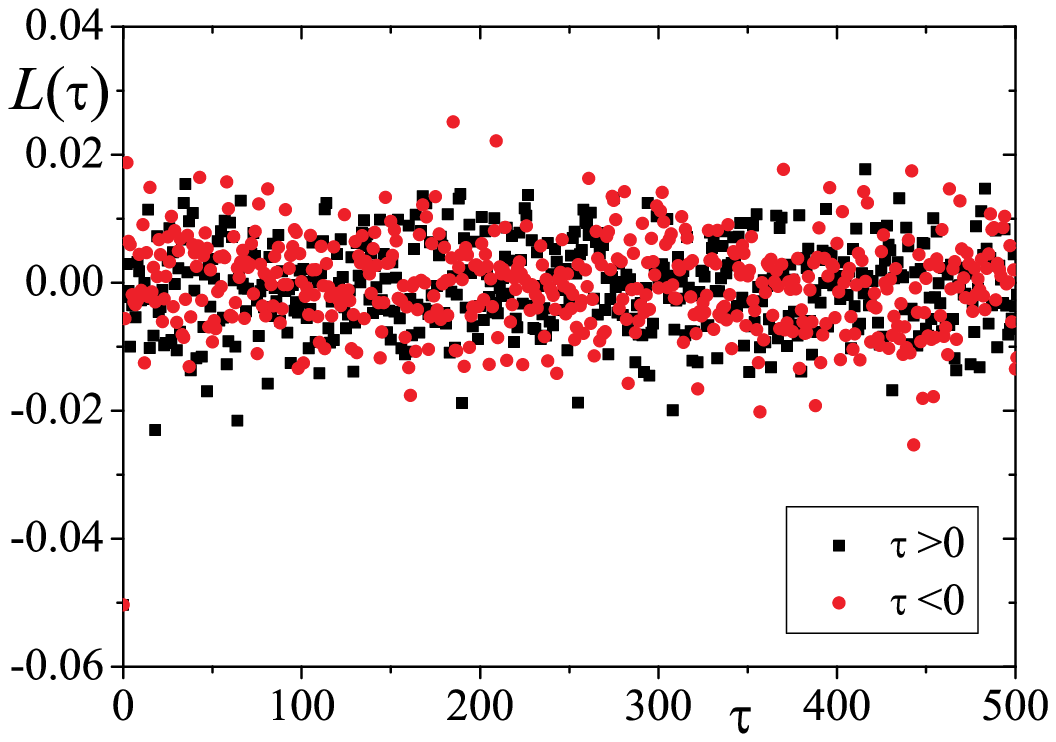} %
\includegraphics[width=0.48\columnwidth,angle=0]{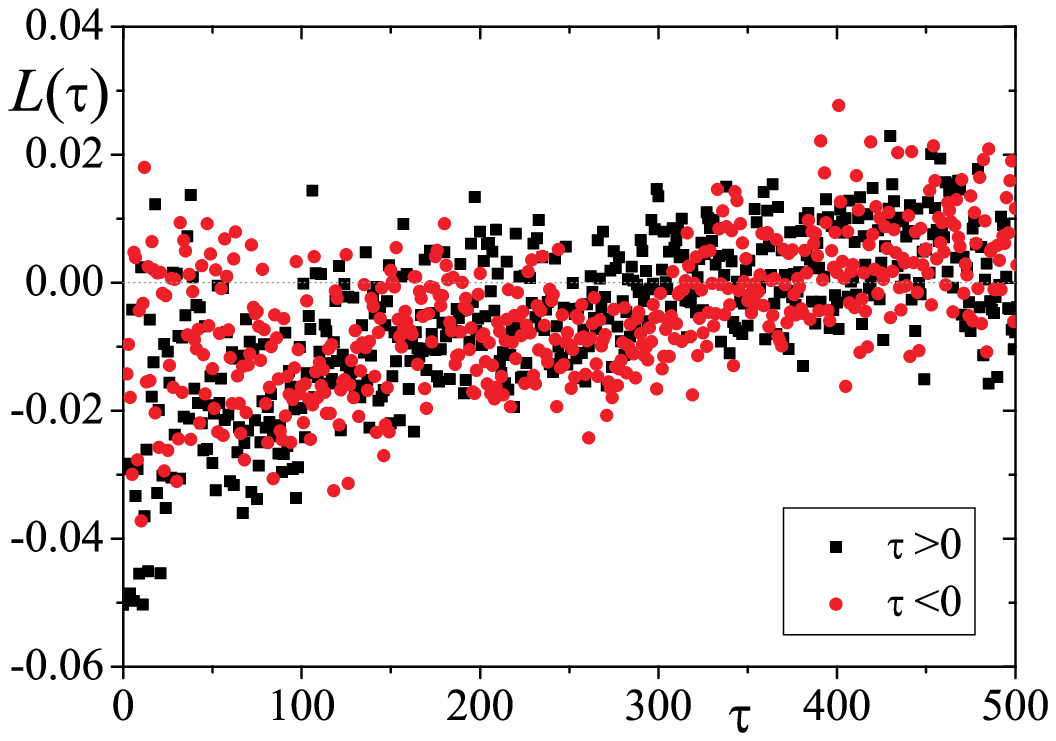} %
\includegraphics[width=0.48\columnwidth,angle=0]{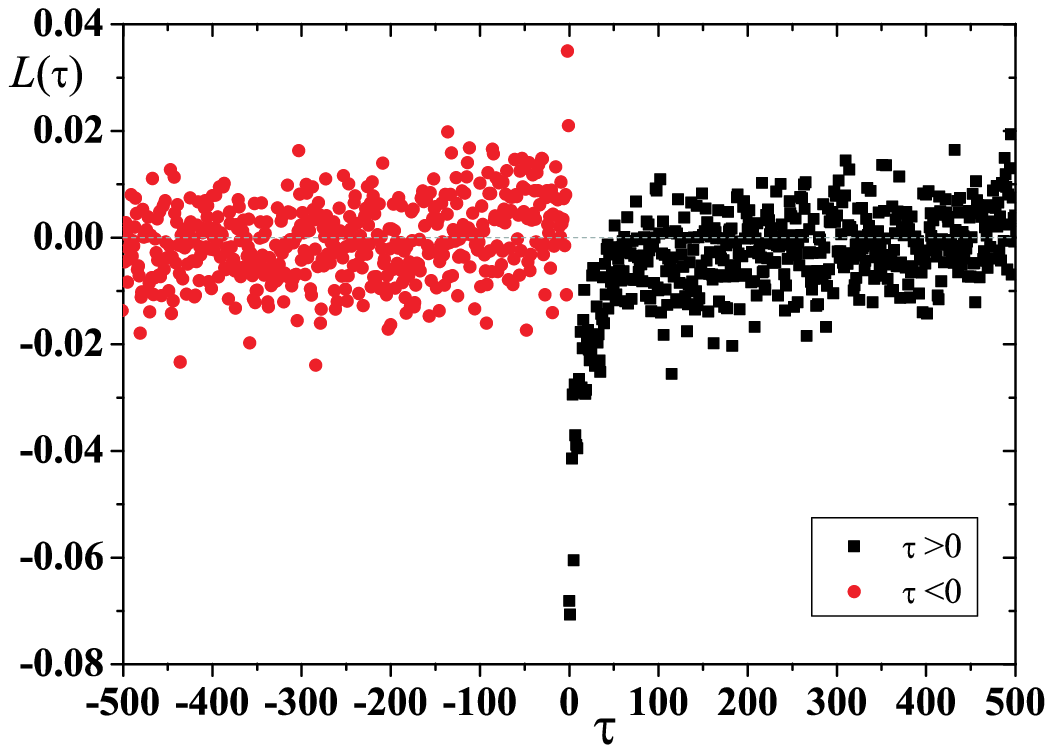}
\end{center}
\caption{(colour on-line) $L\left( \protect\tau \right) $ \textit{vs.} $%
\protect\tau $ of $DTB3$ (left), and shuffled $\left\{ r_{t}\right\} $
(centre left). Comparing both panels and taking into account noise level
(dashed lines) it is visible the existence of a functional form for $L\left(
\protect\tau \right) $. Centre right panel: $L\left( \protect\tau \right) $
for $\protect\tau >0$ and $-L\left( -\protect\tau \right) $ for $\protect%
\tau <0$ \textit{vs.} $\protect\tau $. At right panel, $L\left( \protect\tau %
\right) $ vs. $\protect\tau $ of $SP500$ for mere illustration purposes.}
\label{fig-4}
\end{figure*}

Another property we have analysed are the correlations between fluctuations
and absolute fluctuations~\cite{bouchaud-leverage},%
\begin{equation}
L\left( \tau \right) =\frac{\left\langle r_{t}^{(\prime )}\,\left[ r_{t+\tau
}^{(\prime )}\right] ^{2}\right\rangle }{\left\langle \left[ r_{t+\tau
}^{(\prime )}\right] ^{2}\right\rangle ^{2}}.
\end{equation}%
It has been verified in several securities and financial indices that $%
L\left( \tau \right) =0$ for $\tau <0$, and $L\left( \tau \right) \sim -\exp %
\left[ -\tau /\lambda \right] $ for $\tau \geq 0$. This behaviour, known has 
\textit{leverage effect}~\cite{leverage}, is intimately related to
risk-aversion and negative skew of price fluctuations\ PDF. In defiance of
the noisy $L\left( \tau \right) $ which has inhibited us to present a trusty
quantitative description, it is plausible to affirm that $DTB3$ fluctuations
also show time symmetry breaking, but in an \textit{antisymmetrical fashion}%
, \mbox{$L\left( -\tau \right)=-L\left( \tau \right) $}, as it is
understandable from Fig.~\ref{fig-4}. For $\tau <0$, there is a positive
correlation between fluctuations and future absolute fluctuations, whereas for $%
\tau >0$, there exists a negative correlation between fluctuations and
future absolute fluctuations. This antisymmetric behaviour has clear implications on
dynamical mimicking. As an example, the Heston approach to financial
fluctuations~\cite{heston,yakovenko}, in which the noises of stochastic
equations for the fluctuation and instantaneous variance are
anti-correlated, must be modified in order embrace our empirical
observations of T-bill $\log $-fluctuations.

\medskip

To summarise, in this manuscript we have analysed a set of statistical
properties of daily fluctuations of the $3$-month T-bill trading value, a
highly liquid security. Our results have shown important differences between
this financial instrument and a paradigmatic example of financial securities
statistical properties, the daily fluctuations of $SP500$ index which also
presents similar properties to other debt bonds~\cite{books}. Specifically,
we have verified that T-bill daily fluctuations PDF belong to the $\alpha $%
-stable class of distributions, while other liquid securities that have been
studied so far present the Gaussian distribution as the attractor in PDF
space. This represents a fundamental justification for the well-known
difficulties on the construction (namely specification) and implementation
(namely identification and estimation)\ of generalised spot interest rate
models~\cite{t-bill-problems}, which are always built assuming a finite
standard deviation, unlike L\'{e}vy-Gnedenko class of random variables.
Moreover, we have unveiled that the fluctuations-fluctuations magnitude
correlation function presents an antisymmetric form, \textit{i.e.}, a
different behaviour than the \textquotedblleft leverage
effect\textquotedblright\ that has been verified in other securities.

Our results emphasise the idea that liquidity is not the only factor to take
into account when we aim to define a behavioural class for financial
securities~\cite{matia-amaral-bp,eisler-kertesz-scaling}. Properties such as
the nature of the financial instrument under trading are actually relevant
for its dynamics and categorisation. We address to future work the
development of dynamical scenarios capable of reproducing the statistical
properties we have presented herein.

\bigskip

SMDQ acknowledges C. Tsallis for several discussions on central limit
theorems, L. Borland for practical aspects of financial trading, R. Rebonato
for bibliographic references, and two anonymous colleagues for their comments
that boosted the contents of this manuscript. The work presented benefited
from infrastructural support from PRONEX/MCT (Brazilian agency) and
financial support from FCT/MCES (Portuguese agency).

\end{document}